\documentclass[aps,prc,preprintnumbers,superscriptaddress,twocolumn, showpacs]{revtex4-1}

\usepackage{graphicx}
\usepackage{dcolumn}
\usepackage{bm}
\usepackage{epsf}
\usepackage{epsfig}
\usepackage{amssymb}
\usepackage{verbatim}
\usepackage{amsmath}
\usepackage{longtable}
\usepackage[utf8]{inputenc}
\usepackage{helvet}
\usepackage[polish,english]{}
\usepackage{subdepth}

\begin{document}

\title{Level densities and thermodynamical properties of Pt and Au isotopes}

\author{F.~Giacoppo}
\email[]{francesca.giacoppo@fys.uio.no}
\affiliation{Department of Physics, University of Oslo, N-0316 Oslo, Norway}

\author{F.L.~Bello Garrote}
\affiliation{Department of Physics, University of Oslo, N-0316 Oslo, Norway}

\author{L.A.~Bernstein}
\affiliation{Lawrence Livermore National Laboratory, 7000 East Avenue, Livermore, CA 94550-9234, USA}

\author{D.L.~Bleuel}
\affiliation{Lawrence Livermore National Laboratory, 7000 East Avenue, Livermore, CA 94550-9234, USA}

\author{T.K.~Eriksen}
\affiliation{Department of Physics, University of Oslo, N-0316 Oslo, Norway}

\author{R.B.~Firestone}
\affiliation{Lawrence Berkeley National Laboratory, 1 Cyclotron Road, Berkeley, CA 94720-88R0192, USA}

\author{A.~G\"{o}rgen}
\affiliation{Department of Physics, University of Oslo, N-0316 Oslo, Norway}

\author{M.~Guttormsen}
\affiliation{Department of Physics, University of Oslo, N-0316 Oslo, Norway}

\author{T.W.~Hagen}
\affiliation{Department of Physics, University of Oslo, N-0316 Oslo, Norway}

\author{B.V.~Kheswa}
\affiliation{Department of Physics, University of Oslo, N-0316 Oslo, Norway}
\affiliation{iThemba LABS, P.O. Box 722, 7129 Somerset West, South Africa}

\author{M.~Klintefjord}
\affiliation{Department of Physics, University of Oslo, N-0316 Oslo, Norway}

\author{P.E.~Koehler}
\affiliation{Department of Physics, University of Oslo, N-0316 Oslo, Norway}
\affiliation{Air Force Technical Applications Center, Patrick Air Force Base, Florida, USA}

\author{A.C.~Larsen}
\affiliation{Department of Physics, University of Oslo, N-0316 Oslo, Norway}

\author{H.T.~Nyhus}
\affiliation{Department of Physics, University of Oslo, N-0316 Oslo, Norway}

\author{T.~Renstr{\o}m}
\affiliation{Department of Physics, University of Oslo, N-0316 Oslo, Norway}

\author{E.~Sahin}
\affiliation{Department of Physics, University of Oslo, N-0316 Oslo, Norway}

\author{S.~Siem}
\affiliation{Department of Physics, University of Oslo, N-0316 Oslo, Norway}

\author{T.~Tornyi}
\affiliation{Department of Physics, University of Oslo, N-0316 Oslo, Norway}
\affiliation{Institute of Nuclear Research of the Hungarian Academy of Sciences (MTA Atomki), H-4001 Debrecen, Hungary}

\date{\today}
\begin{abstract}
The nuclear level densities of $^{194-196}$Pt and $^{197,198}$Au below the neutron separation energy have been measured using transfer and scattering reactions. All the level density distributions follow the constant-temperature description. Each group of isotopes is characterized by the same temperature above the energy threshold corresponding to the breaking of the first  Cooper pair. 
A constant entropy excess $\Delta S=1.9$ and $1.1$ $k_B$ is observed in $^{195}$Pt and $^{198}$Au with respect to $^{196}$Pt and $^{197}$Au, respectively, giving information on the available single-particle level space for the last unpaired valence neutron. The breaking of nucleon Cooper pairs is revealed by sequential peaks in the microcanonical caloric curve.
 \end{abstract}

\pacs{21.10.Ma, 27.80.+w, 25.40.Hs, 24.10.Pa}

\maketitle

\section{Introduction\label{intro}}
A detailed knowledge of the nuclear level density is fundamental to understanding reaction mechanisms when the number of levels involved is too large to be treated individually.
Hence, several phenomena in nuclear physics, astrophysics and nuclear reactor science, such as multifragmentation reactions, thermonuclear reaction rates and fusion-fission cross sections, are usually modeled using the level density as a key ingredient.

A wide collection of experimental data, mostly below the particle separation threshold, is currently available for stable and close-to-stability isotopes~\cite{RIPL3, OsloData}. From the theoretical point of view, ever since the seminal work of H. Bethe~\cite{Bethe}, several analytical expressions of the nuclear level density as  function of the excitation energy $E$, the spin $J$ or the angular momentum distribution have been derived. Elaborate microscopic models embody the main effects that substantially influence the density of levels in atomic nuclei, i.e. shell effects, pairing correlations and collective excitations~\cite{MonteCarlo1, MonteCarlo2, HFB}.

In contrast to these microscopic descriptions, simple phenomenological models, such as the back-shifted Fermi gas model~\cite{FermiGas} and the composite Gilbert and Cameron formula~\cite{GilbertCameron}, are usually adopted to globally reproduce the available experimental data, although they lack in a solid theoretical basis. The latter includes a constant-temperature behavior of the level density up to a certain excitation energy where the pairing correlations disappear ($\sim$10 MeV) and then a Fermi-gas formula with an energy shift is applied.

The constant-temperature picture describes well the functional form of the level density in the quasicontinuum region, i.e.~between the discrete levels and the particle separation energy, for heavy deformed nuclei belonging to the rare-earth and actinide series (see Ref.~\cite{MorettoPRL}, Fig.~3 and references therein).

In this paper we investigate for the first time the nuclear level density of $^{194-196}$Pt and $^{197,198}$Au, using the Oslo method~\cite{Schiller, Larsen}. This analytical procedure allows to extract simultaneously the nuclear level density and the $\gamma$-ray strength function from particle-$\gamma$ coincidence measurements. 
Pt and Au isotopes are located in the transitional region between strongly-deformed and spherical nuclear shapes. In particular the structure of $^{196}$Pt has been experimentally established to correspond to a triaxial $\gamma$-soft configuration with a tendency to an oblate shape~\cite{gammasoft_196Pt}.

We can classify the thermal behavior of the above-mentioned nuclear systems in the framework of the microcanonical ensemble, using the experimental level density as the partition function. Therefore, the entropy and other fundamental quantities (temperature and heat capacity) can be extracted to give more insight into the statistical properties of the many-body nuclear system. In particular, fine structures in the entropy distribution as a function of the excitation energy reveal information on the quenching of pairing correlations in atomic nuclei.  These residual interactions lead to effects similar to the superconductivity in metals and are successfully described by applying the Bardeen-Cooper-Schrieffer (BCS) theory~\cite{BCS} to the case of a finite Fermi system such as the nucleus.

\begin{figure*}[t]
\centering
\includegraphics[width=17.9cm]{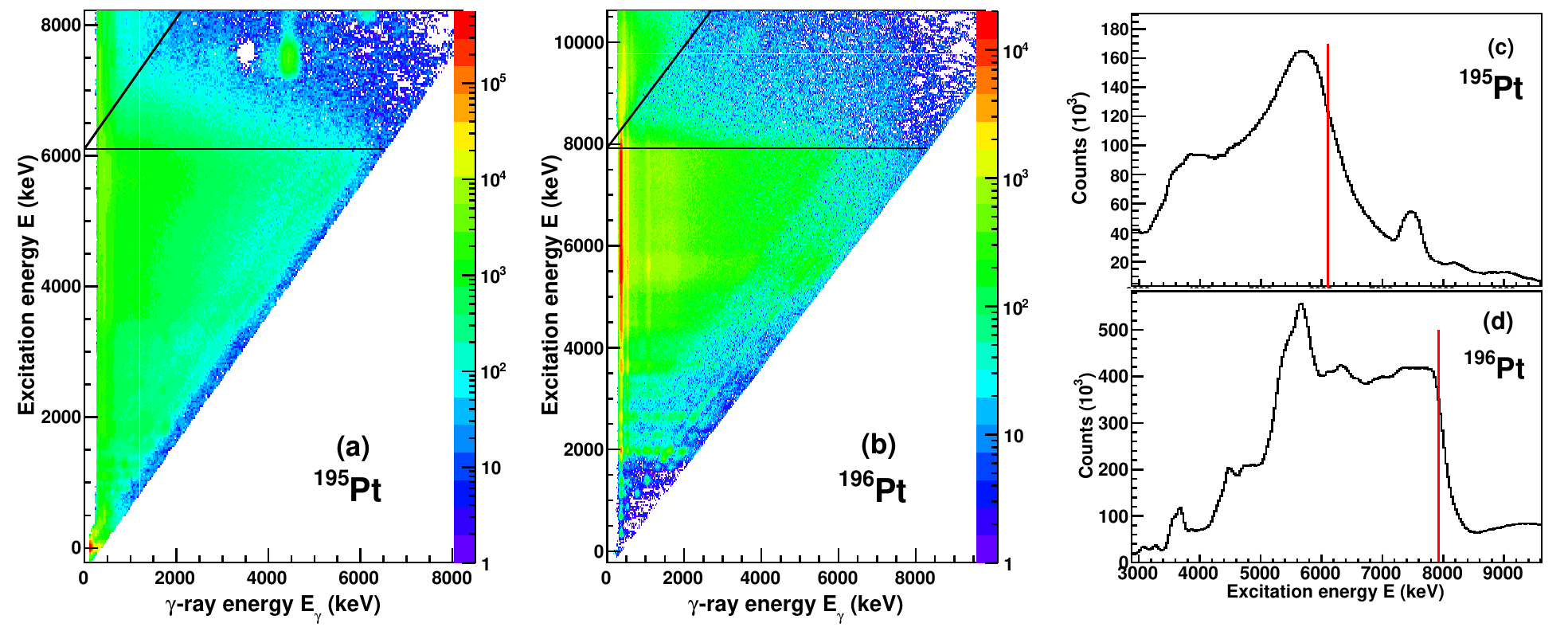}   
\caption{(Color online) The particle-$\gamma$ coincidence matrices for (a) $^{195}$Pt($p,p'$)$^{195}$Pt and (b) $^{195}$Pt($d,p$)$^{196}$Pt after being unfolded with the NaI response functions. The black horizontal lines indicate the neutron separation energy  S$_n$ and the diagonal ones the case of a neutron emitted (and not detected) with zero kinetic energy together with a particle and $\gamma$-rays. The right panel shows the projection along the excitation energy $E$ axis of the two matrices that is (c) the proton and (d) deuteron spectra, respectively. Note how the particle emission drops abruptly above S$_n$ (vertical red line) when an even-mass nucleus is formed ($^{196}$Pt) whereas a less steep decrease is observed in the proton distribution up to $S_{n}$+1.1 MeV. The peak at $E\simeq7.5$~MeV in (a) and (c) is due to the target contamination with $^{12}$C.
 \label{UN_matrices}}
\end{figure*}

In the next Section the details of the experimental technique and the data analysis are described. Sec.~\ref{levelDensity} presents the normalized level density distributions, while the thermodynamics is discussed in Sec.~\ref{thermo}. Finally, concluding remarks are given in Sec.~\ref{conclusions}.      

\section{Experimental setup and\\   data analysis \label{analysis}}

The $^{195}$Pt($d,p$), $^{195}$Pt($p,p'$) and $^{195}$Pt($p,d$) reactions were studied using a self-supporting $^{195}$Pt target enriched to 97.3\% and with a mass thickness of 1.50(15)~mg/cm$^2$. The target was first irradiated for 5 days by a deuteron beam with an energy of 11.3 MeV and an intensity of 1~nA delivered by a MC-35 Scanditronix cyclotron at the Oslo Cyclotron Laboratory (OCL). The second experiment lasted 6 days: this time a 1.8 nA proton beam accelerated to 16.5 MeV was used. For the gold campaign, two identical self-supporting $^{197}$Au targets with a thickness of $\sim 1.93$ mg/cm$^2$ were irradiated with a deuteron beam (12.5 MeV) and a  $^{3}$He beam (34.0 MeV), respectively.

Particle-$\gamma$ coincidences were recorded in the Silicon Ring (SiRi) particle detector system~\cite{SiRi} and the CACTUS multidetector array~\cite{CACTUS}. The former consists of eight trapezoidal Si ${\Delta}$E-E telescopes mounted in a ring at 5 cm distance from the target. SiRi was placed in backward direction with respect to the beam direction, in order to minimize the detection of projectiles that undergo elastic scattering on the target. It has a solid-angle coverage of $\sim6\%$ of 4$\pi$. Each telescope consists of a thin (130 ${\mu}$m) front detector segmented into 8 strips and a 1550 ${\mu}$m thick back detector for a total of 64 independent telescopes covering eight scattering angles between  $\theta=126^{o}$ and $\theta=140^{o}$, with a resolution of $\Delta\theta=2^{o}$. By plotting the energy deposited into the ${\Delta}$E detector versus the E detector the different types of charged particles are uniquely identified. From the reaction kinematics and Q-value, the energy and angle of the emitted particle, one can extract the excitation energy $E$ of the residual nucleus. A resolution of $\Delta E\sim100$, 150 and 200 keV has been reached for the $p$-, $d$- and $^3$He-induced reactions, respectively. CACTUS is a spherical array of 26 collimated $5''\times5''$ NaI(Tl) $\gamma$-ray detectors with a total solid-angle coverage of 16.2\% out of 4$\pi$, surrounding the target point, and with a total detection efficiency of 14.1(1)$\%$ at $E_{\gamma}$= 1332~keV.

The ejectile-$\gamma$ coincidences are recorded on an event-by-event basis. The time resolution is $\Delta t\approx$15 ns.
The spectrum of emitted $\gamma$-rays can then be analyzed for a given excitation energy of the residual
nucleus after being corrected for the known CACTUS response function through an unfolding procedure~\cite{Unfolding}. In this work a recent version of the NaI response function has been used: the relative efficiency as a function of the $\gamma$ energy has been reliably extracted for several $\gamma$ lines from excited states in $^{13}$C, $^{17}$O, $^{28}$Si and $^{56,57}$Fe. Ultimately, the unfolding method provides Compton background-subtracted  $\gamma$-ray spectra with unaltered statistical fluctuations.

In Fig.~\ref{UN_matrices} the matrices of unfolded $\gamma$ spectra for each excitation energy bin of $^{195,196}$Pt are shown. The triangles below the neutron separation energy $S_{n}$ correspond to the particle-$\gamma$ coincidences from the ($d,p\gamma$)$^{196}$Pt and ($p,p'\gamma$)$^{195}$Pt reactions, respectively. One can see that for the even-mass isotope the emission of $\gamma$-rays drops suddenly above S$_n$ where the neutron channel ($d,pn\gamma$)$^{195}$Pt is open. In contrast, for $^{195}$Pt a significant amount of $\gamma$-rays is emitted for energies up to $E\simeq S_n+1.1$ MeV. This distinct behavior can be clearly seen if the projection on the $E$-axis is taken (see Fig.~\ref{UN_matrices}-right panel). To explain the two diverse features one should look at the levels populated in the neutron channels ($p,p'n\gamma$)$^{194}$Pt and ($d,pn\gamma$)$^{195}$Pt. In the latter case, since the final nucleus has an odd number of nucleons, it has several levels with spins ranging over a broad distribution, even at low energy. In the case of $^{194}$Pt only few levels with $J^{\pi}$=$0^+$, $2^+$, $3^+$and $4^+$ are available. 
Then, when higher angular momenta are transferred during the reaction, states above $S_n$ of the compound nucleus $^{195}$Pt are populated and decay by $\gamma$ emission.

One of the main components of the Oslo method is an iterative subtraction technique developed to separate out the distribution of primary (first-generation) $\gamma$ transitions from the cascade of $\gamma$-rays originating from states at a given excitation energy~\cite{Primarygen}. The basic assumption of this technique is the independence of the $\gamma$-decay pattern from the way the states are populated, i.e.~directly by a nuclear reaction or as part of a de-excitation cascade. This assumption is valid for levels fed with comparable probability by the two processes. It is also valid in the region of high level density where the nucleus seems to thermalize in a compound-like phase before $\gamma$ emission. In our analysis we consider initial excitation energy bins containing many levels.
Hence, the corresponding $\gamma$-ray spectra are on average independent of the population path.

\begin{figure}[ht]
\includegraphics[width=\columnwidth, height=15.4cm]{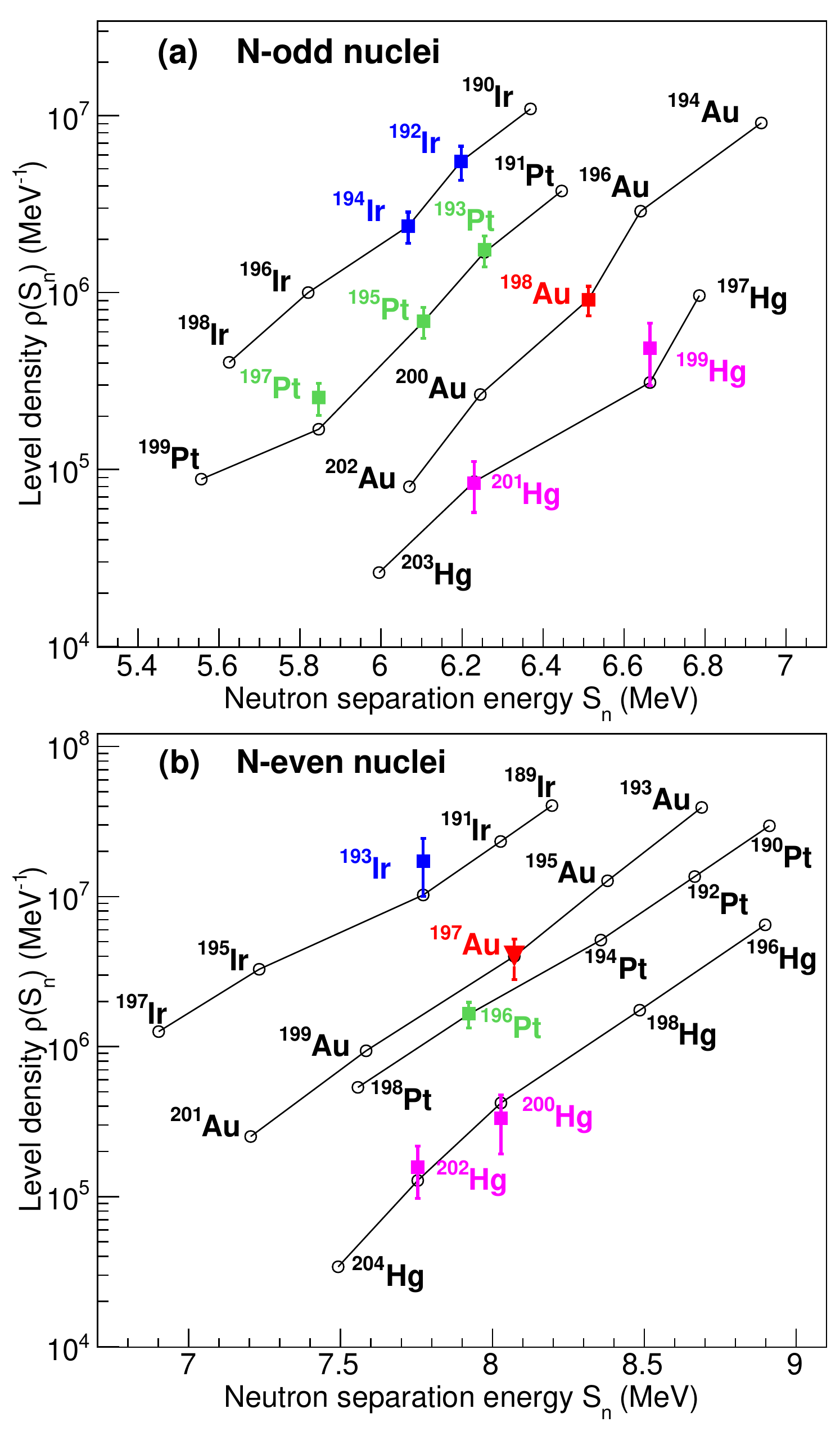}
\caption{(Color online) Estimation of $\rho(S_n)$ for $^{197}$Au. The total level density at the neutron separation energy is calculated from experimental value of $D_0$~\cite{RIPL3, Paul_Pt_astro} for neighboring $N$-odd and $N$-even isotopes (colored filled squares). (a) For $^{198}$Au and other $N$-odd nuclei a comparison with the systematics (open circles and black line) is done~\cite{EB2009}. A correction factor of $1.6-1.85$ is applied for each isotopic chain. (b) The same factor is used for the $N$-even isotopes. The red triangle is the estimated value of $\rho(S_n)$ for $^{197}$Au. \label{Sn_Rho}}
\end{figure}
\begin{table*}[th]  
 \caption{Input parameters for the normalization of the level density of $^{194-196}$Pt and $^{197,198}$Au.\label{nld_models}}
\begin{ruledtabular}
 \begin{tabular}{l*{8}{c}}
\renewcommand{\thefootnote}{\thempfootnote}
    		    &\multicolumn{4}{c}{} &\multicolumn{2}{c}{CT model~\cite{GilbertCameron}} &\multicolumn{2}{c}{HFB+comb model~\cite{HFB}}\\	
Nucleus   &   $S_n$ [MeV]   &   $D_0$ [eV]   & $\sigma$(S$_n$)    &   $\rho(S_n)$  [10$^4$ MeV$^{-1}]$   &   $T_{CT}$ [MeV]   &   $E_0$ [MeV] &$\mathfrak{a}$&$\delta $[MeV] \\
\hline\\
$^{194}$Pt     &8.357       & -        		&  5.06(51)    & 512(154)\footnote{Estimated from systematics~\cite{EB2009, RIPL3} as shown in Fig.~\ref{Sn_Rho}.\label{footnote_tab1}}     	&0.63       & -1.08     &-0.20     &-0.08    \\
$^{195}$Pt     &6.105       & 71.8(29)      &  4.92(49)     & 69(14)     	&0.63       & -2.07     &-0.44      &-0.32    \\
$^{196}$Pt     &7.922       & 15.93(41)    &  5.04(50)     & 165(32)     	&0.63       & -0.81     &-0.35      &0.38      \\
$^{197}$Au   &8.072       & -                  & 5.15(51)      &400(120)\footnotemark[\value{mpfootnote}]   &0.68       & -2.0       &-0.32      &0.06      \\
$^{198}$Au   &6.512       & 15.5(8)     & 5.08(51)      &91(17)      &0.67       & -2.42     &-0.41      &-0.30    \\
\end{tabular}
\end{ruledtabular}
 \end{table*}
\begin{figure*}[t]
\includegraphics[width=2\columnwidth]{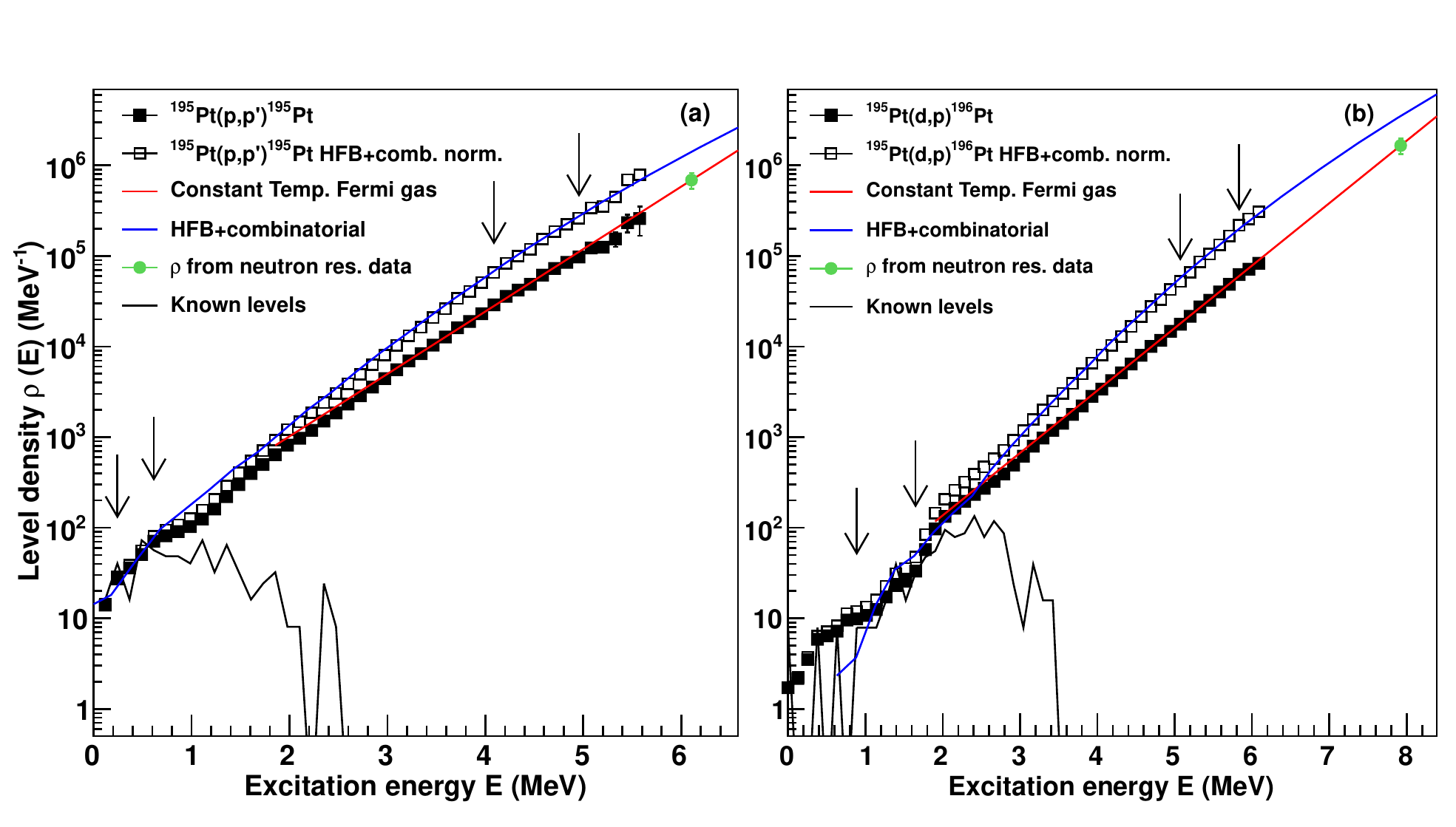}
\caption{(Color online) Normalization of the level density (black filled squares) of (a) $^{195}$Pt and (b) $^{196}$Pt to discrete levels at low energy (black line) and to $\rho(S_n)$ at high energy (green filled circle). The vertical arrows define the region where the normalization has been applied. The trend of the experimental distribution is compatible with the CT approach (red line). Another normalization (black open squares), following the HFB plus combinatorial approach (blue line), is proposed, see the text.\label{NLD_195Pt_196Pt}}
\end{figure*}
From the matrix of first generation $\gamma$-rays tagged in excitation energy, $P(E, E_{\gamma})$, the functional form of the nuclear level density $\rho$ and the $\gamma$ transmission coefficient $\mathcal{T}$ can be derived through a simultaneous fit, according to the following factorization:
\begin{equation}
P(E, E_{\gamma})\propto\rho(E-E_{\gamma})\mathcal{T}(E_{\gamma})
 \label{eq-1}
\end{equation}
where $\rho(E-E_{\gamma})$ is the level density at the final excitation energy $E_{f}=E-E_{\gamma}$. This equation is in accordance with the Fermi's golden rule~\cite{Fermi}: the decay probability is proportional to the level density at the final state and the squared transition matrix element between the initial and the final state. The decomposition of $P(E, E_{\gamma})$ into two independent functions of $E_{f}$ and $E_{\gamma}$ is justified in the limit of a statistical decay  process, after  a compound state is formed during the reaction~\cite{BohrMottelson}. A window in initial excitation energy is then selected where compound state formation is predominant: the lower limit is at $\sim2.0-4.0$ MeV for the Pt and Au isotopes presented in this work.
Above the neutron separation energy the final reaction channel is ambiguous since the compound state may evaporate neutrons (which are not detected in these experiments) before $\gamma$ decay. The transmission coefficient is independent of the excitation energy according to the Brink-Axel hypothesis~\cite{Brink, Axel} in its generalized form: collective modes built on any excited state have the same properties as the ones built on the ground state. This hypothesis is not valid for reactions where high temperatures and spins are involved. Since in the present cases the nuclear temperature is relatively low, as it will be shown in the following, and the spin distribution is centered at $J\approx4-5$, significant deviations from the mentioned assumptions are not expected.

Relation~(\ref{eq-1}) has an infinite number of possible solutions generated by the transformations:
\begin{align}
\tilde{\rho}(E-E_{\gamma}) & =A\mathrm{e}^{{\alpha}(E-E_{\gamma})}\rho(E-E_{\gamma})
 \label{eq-2}\\
\tilde{\mathcal{T}}(E_{\gamma}) & =B\mathrm{e}^{{\alpha}E_{\gamma}}\mathcal{T}(E_{\gamma})
 \label{eq-3} .
\end{align}
To obtain the absolute level density and $\gamma$ transmission coefficient a set of parameters $A$, $B$ and $\alpha$ has to be determined using independent experimental data. Finally, the $\gamma$ transmission coefficient is associated to the $\gamma$-ray strength function $f(E_{\gamma})$ by the relation $\mathcal{T}(E_{\gamma})\propto\sum E_{\gamma}^{2L+1}f_{XL}(E_{\gamma})$ where $X$ and $L$ stand for the electromagnetic character and the multipolarity of the $\gamma$-ray, respectively.

In the next section, details of the normalization procedure will be presented and discussed with special focus on the level density distribution.

\section{Nuclear Level Densities\label{levelDensity}}

\begin{figure*}[t]
\includegraphics[width=2\columnwidth]{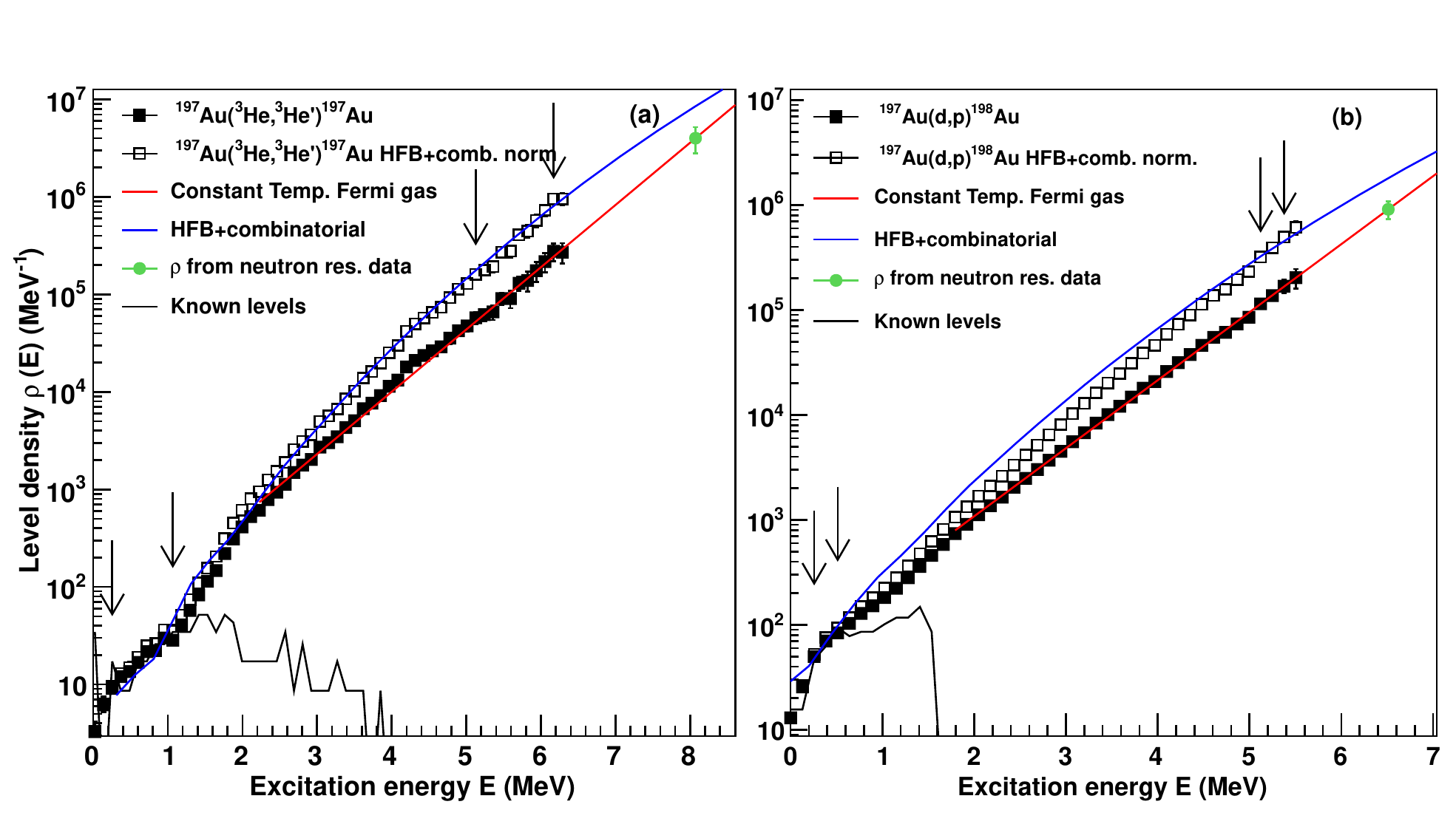}
\caption{(Color online) Level density of (a) $^{197}$Au and (b) $^{198}$Au. The black filled squares are normalized to discrete levels at low energy (black line) and to $\rho(S_n)$ at high energy (green filled circle) with the constant temperature extrapolation (red line). The same data are also normalized to the HFB plus combinatorial approach (blue line) and are represented by black open squares.\label{NLD_197Au_198Au}}
\end{figure*}

To normalize the level density the amplitude $A$ and the slope $\alpha$ in Eq.~(\ref{eq-2}) are extracted from a fit to known densities at low and high excitation energies. In the former case the discrete levels with energy $E<3.9$~MeV are taken from literature~\cite{NNDC}. The experimental level scheme is far from complete when a density of about 50-100 levels per MeV is reached.
At the neutron separation energy $S_{n}$ the total level density $\rho(S_{n})$ is calculated from measured values of the neutron resonance spacing $D_0$ (s-wave) in the corresponding ($n,\gamma$) reaction on the A-1 target nucleus~\cite{Schiller, Larsen}. In order to obtain $\rho(S_{n})$, the ground state spin $I_t$ of the target nucleus and the spin distribution at $S_{n}$ must be taken into account.
As suggested in Ref.~\cite{Ericson}, the spin distribution is generally expressed with a Gaussian-like formula containing a single free parameter, the spin cutoff  $\sigma$:
\begin{equation}
g(J,\sigma)\simeq\frac{2J+1}{2\sigma^2}\textit{e}^{-\frac{J(J+1)}{2\sigma^{2}}}
\label{eq-spin} .
\end{equation}
In this work we adopt the empirical expression of the dependence of $\sigma$ on mass and excitation energy proposed by von Egidy and Bucurescu in their systematic study of level density parameters~\cite{EB2009}:
\begin{equation}
\sigma^2=0.391A^{0.675}E'^{0.312}
\end{equation}
where A is the mass number and $E'=E-0.5Pa'$ with $Pa'$ being the deuteron pairing energy.
The parity distribution at $S_{n}$ is assumed to be symmetric, as supported by theoretical microscopic calculations for the nuclei under study~\cite{HFB}.

Very recent and accurate measurements of the level spacing $D_0$ of s-wave neutrons are available for Pt isotopes~\cite{Paul_Pt_astro}. For $^{198}$Au we adopt the value reported in the Reference Input Parameter Library (RIPL-3) compilation~\cite{RIPL3}. Unfortunately, there exists no information on the neutron level spacing of $^{197}$Au through the ($n,\gamma$) reaction in literature, since  $^{196}$Au is unstable. For this nuclide we estimate $\rho(S_{n})$ from a comparison of the experimental values in neighboring isotopes and the systematics of Ref.~\cite{EB2009}. In Fig.~\ref{Sn_Rho}-(a) the total level density at $S_n$ of $N$-odd $_{77}$Ir, $_{78}$Pt, $_{79}$Au and $_{80}$Hg isotopes is reported: the colored markers refer to $\rho(S_{n})$ values calculated from $D_0$~\cite{RIPL3}. The open circles connected with black lines represent the global systematics~\cite{EB2009}. A scaling factor ranging between 1.6 and 1.85 is applied to reproduce the experimental $\rho(S_n)$ in the $N$-odd nuclei  and the same value is kept for the corresponding $N$-even isotopes (Fig.~\ref{Sn_Rho}-(b)): the agreement is good except for $^{193}$Ir. The uncertainty of $\rho(S_n)$ estimated for $^{197}$Au with this procedure is about 30\%. It is estimated taking into account the uncertainty for $^{198}$Au and the scaling correction to adjust with the systematics. The general trend of $\rho(S_n)$  as a function of $S_n$ is decreasing towards more neutron-abundant isotopes. The input parameters for the normalization of $^{195,196}$Pt and $^{197,198}$Au level densities are listed in Table~\ref{nld_models}.

Figures~\ref{NLD_195Pt_196Pt} and~\ref{NLD_197Au_198Au} show the extracted level densities normalized according to the procedure depicted above. One can see that an extrapolation is needed to connect the level density data points at the highest excitation energies with $\rho(S_n)$: we use the constant temperature (CT) formula~\cite{Ericson, GilbertCameron}
\begin{equation}
\rho_{CT}(E)=\frac{1}{T_{CT}}\textit{e}^{\frac{E-E_{0}}{T_{CT}}}
\end{equation}
that reproduces well the exponential logarithmic trend of the extracted level densities. The values of the temperature $T_{CT}$ and the energy shift $E_0$ are listed in Table~\ref{nld_models}. They are in good agreement with the systematics~\cite{EB2009} and are determined in order to reproduce the experimental value at $S_n$.

\begin{figure}[t]
\includegraphics[width=\columnwidth]{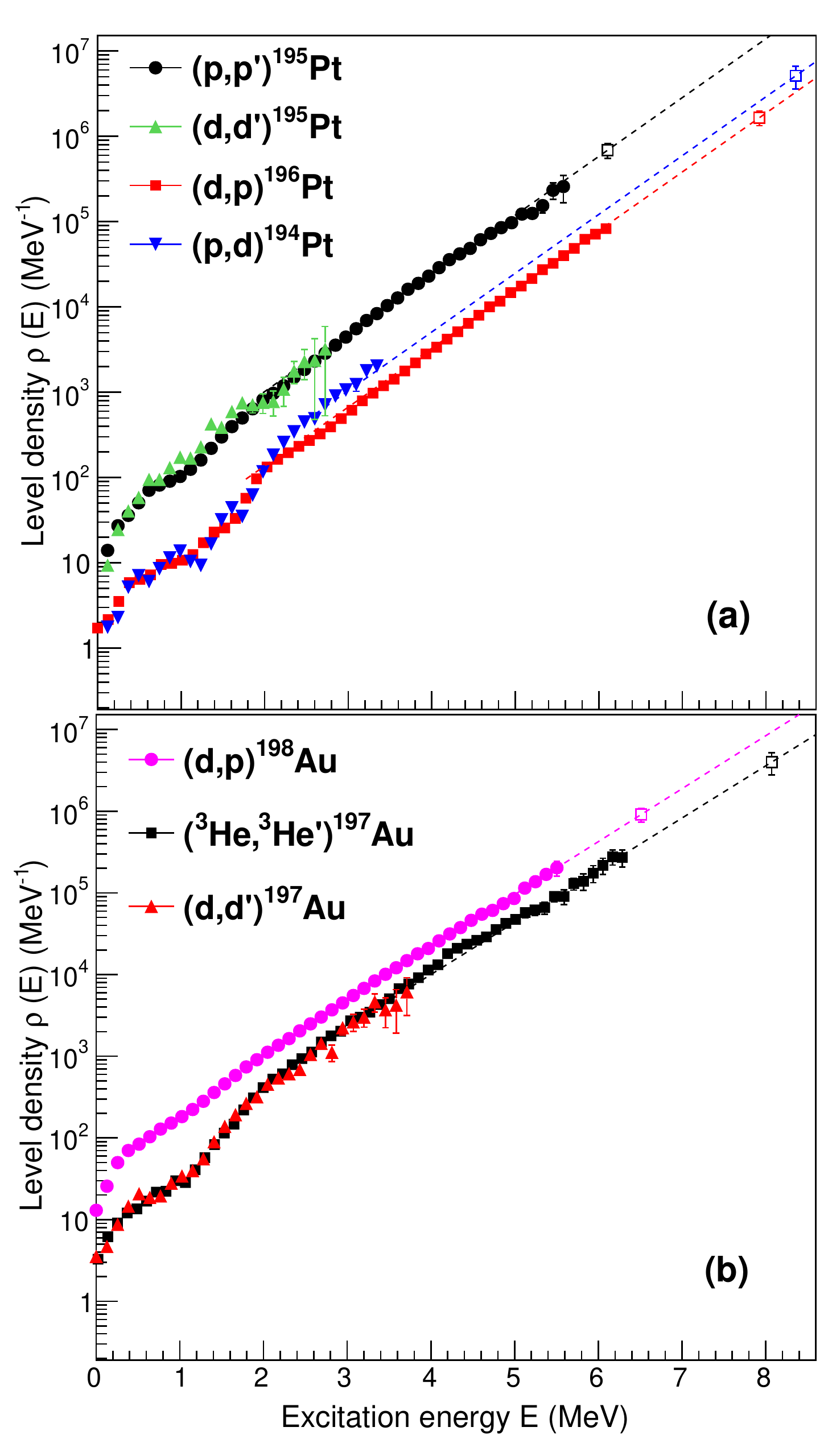}
\caption{(Color online) Comparison of the level density of (a) $^{194-196}$Pt and (b) $^{197}$Au and $^{198}$Au. Both distribution sets follow a straight-line trend in the log plot corresponding to a constant temperature $T_{CT}=0.63$ and $0.68-0.67$~MeV for $^{195,196}$Pt and $^{197,198}$Au, respectively. The staggering of the even-even and even-odd Pt isotopes is much more pronounced than for the odd-even and odd-odd Au nuclei. For $^{195}$Pt and $^{197}$Au data from the ($d,d'$) reaction are also shown.
\label{NLD_Pt_Au_comp}}
\end{figure}

Besides the constant temperature term of the composite phenomenological model, the microscopic combinatorial approach of Goriely {\em et al.}~\cite{HFB} is also included in the comparison. The Hartree-Fock-Bogoliubov (HFB) plus combinatorial method provides a parity as well as energy and spin dependent level density that reproduces fairly well the low-energy discrete  region of our data  and estimates the neutron resonance spacing $D_0$ with a good degree of accuracy.
As described in Ref.~\cite{HFB} the calculations are normalized with the $\delta$ and $\mathfrak{a}$ parameters listed in Table~\ref{nld_models} to be compared with the experimental data. 
This model predicts a more or less pronounced concave curvature of the level density below the particle threshold in contrast with the straight-line behavior of the experimental data in a logarithmic scale, especially for $^{196}$Pt and $^{198}$Au. In addition, the normalization using the microscopical model gives a slightly steeper function with respect to the phenomenological CT model. We can translate this effect as a broader spin distribution accounted by the microscopical model for the nuclei under study. 

Goriely and his coworkers have recently published a new version of the combinatorial model where the collective effects are predicted by a newly derived Gogny interaction~\cite{HFB-Gogny}. For the Pt and Au mass region an increased curvature is expected in the level density below $S_n$: a trend not seen in our data. For this reason a comparison between this model and the experimental data is not shown in this work. A better agreement with the previous formulation of the model is achieved.

In conclusion, the four data sets are best represented by the constant temperature formula above $E\sim2$~MeV.
A remarkable feature of our results is the common parallel trend of the level densities in a log plot (see Fig.~\ref{NLD_Pt_Au_comp}): as expected the odd-mass Pt isotope has a higher level density with respect of the neighbor even-nucleus, corresponding to a scaling factor of 9. For both Pt nuclei we observe a steep increase in the density of levels for $E\leqslant1$~MeV. In particular, the $\gamma$-band in  $^{196}$Pt opens above the $2^{+}$ level at $E=355.7$~keV in the ground band~\cite{gammasoft_196Pt}. 
A second step-like increment is observed between $1-2$~MeV in $^{196}$Pt where the breaking of nucleon Cooper pairs occurs.
This effect will be further discussed in the next section.

In the case of gold (Fig.~\ref{NLD_Pt_Au_comp}-lower panel), the odd-odd isotope shows a higher density of levels with respect to the even-odd neighbor. Again the two distributions are parallel in a log scale but this time the spacing between them corresponds to a scaling factor of only $\sim2.5$.
The level density of $^{197}$Au  has a rapid increase up to $E\sim1.0$~MeV. From 1.0 to 2.0 MeV the slope of the level density becomes even steeper, due to the breaking of Cooper pairs and the availability of more quasiparticles that can combine in all possible configurations. In $^{198}$Au, the density of levels is rapidly increasing after a few hundred keV of excitation energy. The presence of an unpaired neutron and proton, on average, smears out the effect of the first Cooper pair breaking, therefore the level density appears smoothly increasing without any pronounced step-like structure.

Fig.~\ref{NLD_Pt_Au_comp} displays also the level density of $^{195}$Pt and $^{197}$Au extracted from the $^{195}$Pt($d,d'$) and $^{197}$Au($d,d'$) reactions, respectively. The comparison confirms that similar results are obtained for different incident projectiles. Selecting the $^{195}$Pt($p,d$) reaction channel it was also possible to extract the level density of $^{194}$Pt up to $E=3.4$~MeV, using the parameters reported in Table~\ref{nld_models}. The constant temperature extrapolation is calculated with $T_{CT}=0.63$~MeV as for the others Pt isotopes. The functional form of $\rho(E)$ is similar to the one of $^{196}$Pt with an abrupt increase at about $2.0$~MeV. 
\begin{table}[t]  
 \caption{Neutron and proton pairing gap parameters of $^{195,196}$Pt and $^{197,198}$Au~\cite{BohrMottelson,AME2012}.\label{pairing_gap}}
\begin{ruledtabular}
 \begin{tabular}{c*{4}{c}}
Pairing gap &     &   &    & \\
(MeV) &   $^{195}$Pt  &   $^{196}$Pt  & $^{197}$Au   &  $^{198}$Au\\   
\hline\\
$\Delta_n$    &1.02&0.97&0.75&0.66 \\
$\Delta_p$    &0.76&1.04&0.94&0.66\\

\end{tabular}
\end{ruledtabular}
 \end{table}
\begin{figure*}[t]
\includegraphics[width=2\columnwidth]{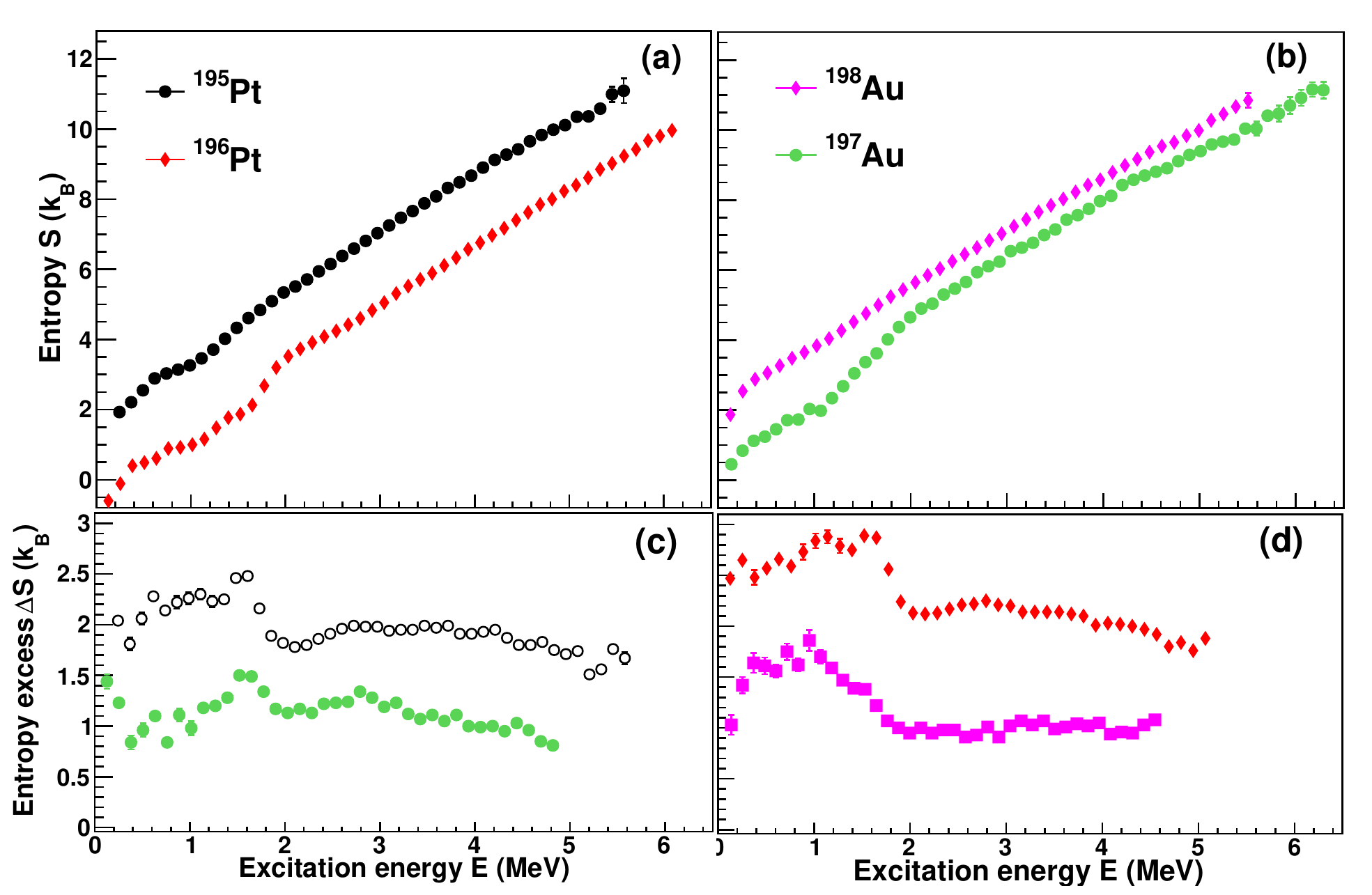}
\caption{(Color online) Entropy curves for (a) $^{195,196}$Pt and (b) $^{197,198}$Au. The lower panels show the entropy differences of (c) $^{195}$Pt (black open circles) and $^{197}$Au (green filled circles) relative to $^{196}$Pt and (d) $^{198}$Au relative to $^{197}$Au (violet filled squares) and $^{196}$Pt (red filled diamonds). The error bars correspond to the propagated error of $\rho(E)$.\label{entropy_Pt_Au}}
\end{figure*}

Another important observation can be made when one compares the level density of neighboring nuclei: as already shown in the actinides, pairs of even-even and odd-even close isotopes have a level density characterized by the same slope, i.e.~the same temperature~\cite{MorettoPRL, actinidesMagne}. This feature is valid also in this case, for soft-deformed nuclei.
If we measure the shift $\Delta$ along the excitation energy axis between $^{195,196}$Pt and $^{197,198}$Au pairs  we obtain a value equal to 1.25 and 0.54 MeV, respectively. The experimental shifts are in qualitative agreement with the neutron pairing gaps $\Delta_{n}$ reported in Table~\ref{pairing_gap}. The latter lists both $\Delta_{n}$ and $\Delta_{p}$ for the four nuclei under study: the neutron and proton pairing gap parameters are determined using the three-point mass-difference formula of Ref.~\cite{BohrMottelson}, from the empirical masses of Pt and Au isotopic and isobaric chains, respectively~\cite{AME2012}.

In the next Section we will focus on the the most complete dataset of $^{195,196}$Pt and $^{197,198}$Au for the extraction of the thermodynamical properties.

\section{ Thermodynamics \label{thermo}}
\begin{figure*}[t]
\includegraphics[width=2\columnwidth]{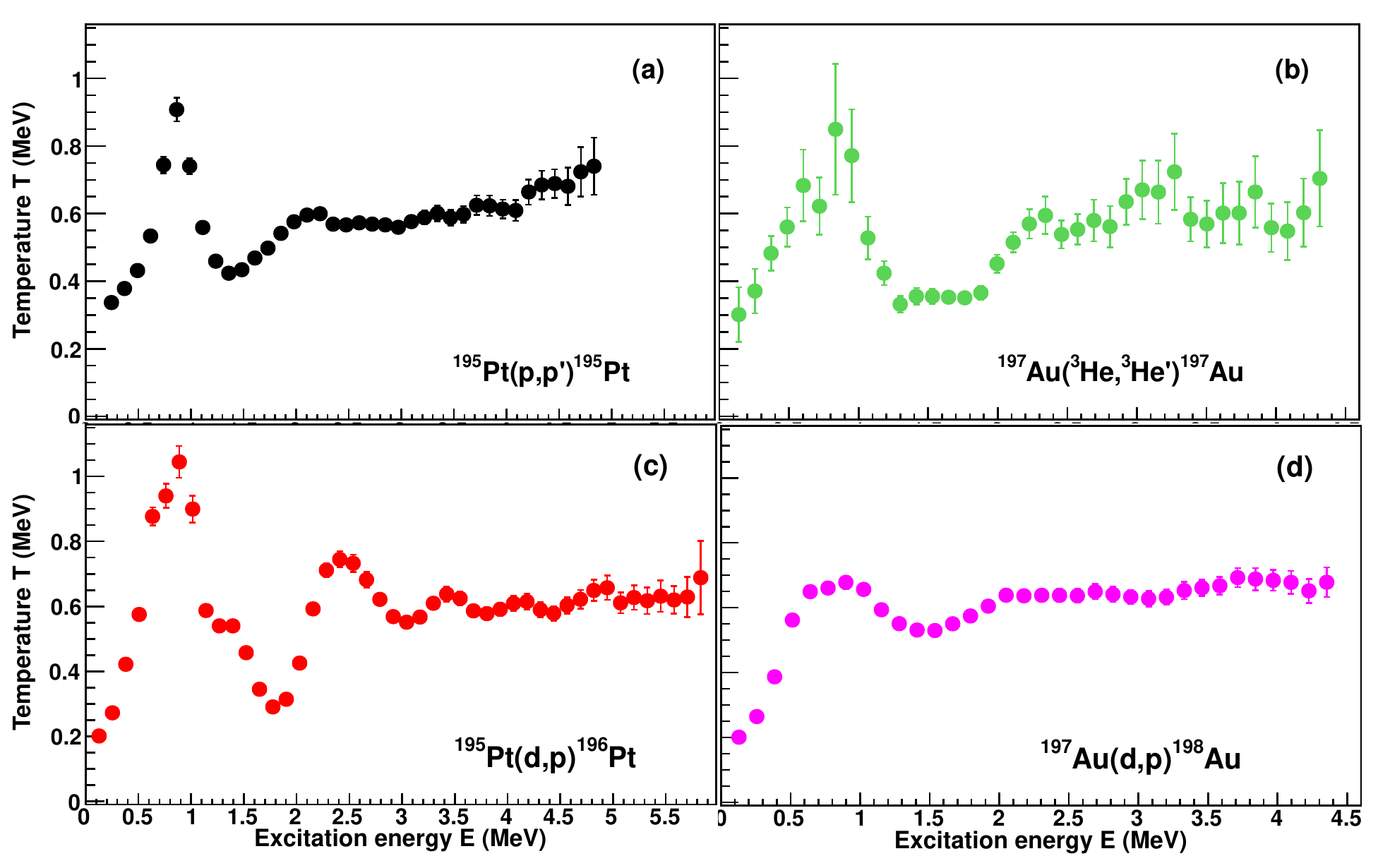}
\caption{(Color online) Experimental temperatures of (a) $^{195}$Pt, (b) $^{197}$Au, (c) $^{196}$Pt and (d) $^{198}$Au as a function of the excitation energy within the formalism of the microcanonical ensemble.\label{temperature_Pt_Au}}
\end{figure*}
An atomic nucleus can be treated as an isolated system with fixed energy and particle number, since the strong force has a short range and the excitation energy is in general not exchanged with the surroundings. Hence, a thermodynamical description of excited nuclei can be derived from their measured level density.

According to these facts, the nucleus can be studied in the framework of the microcanonical ensemble: its level density can be used to define a partition function and the entropy is expressed, according to the Boltzmann's principle, as
\begin{equation}
S(E)=k_{B}\ln W(E)
\end{equation}
where $k_B$ is the Boltzmann constant and $W(E)\propto(2J+1)\rho(E)\delta E$. Here the number of accessible states for the system $W(E)$ is proportional to the density of states, i.e.~to the experimental level density $\rho(E)$ multiplied by the spin degeneracy of magnetic substates. The present experimental technique provides no information about the spins $J$ populated during the reaction and we therefore extract a "reduced" entropy
\begin{equation}
S(E)=\ln\Big[\frac{\rho(E)}{\rho_{0}}\Big]
\end{equation} 
in units of $k_B$, where $\rho_{0}$ is a normalization factor that ensures the validity of the third law of thermodynamics, i.e.~the entropy approaches a constant value at temperatures close to zero. Since the ground state band of an even-even nucleus such as $^{196}$Pt has $W(E)\sim1$ within experimental energy bins of $\sim150$ keV, and it represents an ordered system with all its nucleons paired, its entropy would be zero in the ground state. Therefore $\ln(\rho_{0})$ is set to $-1.37$ $k_B$ and this value is used also for the other nuclei under study. 

The entropies of $^{195,196}$Pt (left) and $^{197,198}$Au (right) are displayed in the upper panels of Fig.~\ref{entropy_Pt_Au}. As already noticed for $\rho(E)$, the entropy curves $S(E)$ are rather linear, however small bumps are visible.  
In $^{196}$Pt a rapid increase of $S(E)$ from 1.9 to 3.9 $k_B$ is observed around 1.9 MeV corresponding to the breaking of a Cooper pair; this step-like increment is expected to occur at about twice the proton or neutron pairing gap. In this even-even nucleus the two pairing gaps $\Delta_{p}$ and $\Delta_{n}$ have similar values, see Table~\ref{pairing_gap}. As a result, the breaking of both a proton and a neutron pair, on average, contributes to such an abrupt increase in the disorder of the nuclear system. The other Pt isotope has already an unpaired neutron that smears out this effect causing the entropy  to increase in a more uniform way and a less pronounced step is visible.

The two Au isotopes form an odd-even and an odd-odd pair: for $^{198}$Au the entropy is high even at low excitation energies and increases linearly. 
In $^{197}$Au the thresholds for breaking a neutron and a proton pair are shifted at about 1.5 and 1.9 MeV, respectively: the combined effect gives a smooth increment in the entropy curve between 1.0 and 2.0 MeV instead of a sharp edge.

The lower panels of Fig.~\ref{entropy_Pt_Au} show the entropy difference $\Delta S$ between the odd and the even mass isotopes. Above 1.8 MeV the entropy excess stabilizes around a value of 2.0 $k_B$ and 1.1 $k_B$ for the Pt and Au pairs, respectively. Accordingly, in $^{195}$Pt an unpaired neutron contributes to the system with an increment in entropy equal to 2.0 $k_B$, corresponding to $\exp(\Delta S)=\exp(2.0)\approx7$ available states per quasiparticle. However, the unpaired nucleon is not identified with a neutron in a specific orbital configuration, it retains the average properties of all the valence orbits under the Fermi surface. At the same time, the extra unpaired neutron in $^{198}$Au has a reduced number of accessible states, giving a contribution to the entropy of the system of only 1.1 $k_B$. Moreover, the entropy excess of $^{197}$Au with respect to $^{196}$Pt is also displayed in the same figure to assess the average contribution of a proton to the entropy of the system. In this case $\Delta S=1.1$ $k_B$: this means that the entropy of the system is largely affected by the neutron configurations, to the same extent as in well deformed nuclei such as actinides or lanthanides~\cite{actinidesMagne,DyMagne}. The valence proton has a reduced phase space, that one would ascribe to the proximity of the shell gap at $Z=82$ that hinders the formation of pairing correlations. However, from the experimental level density of mid-shell actinides, the odd-odd $^{238}$Np~\cite{NpTamas} and the even-odd $^{237}$U~\cite{actinidesMagne}, the deduced entropy excess due to the unpaired proton is also $\Delta S\approx1.1$~$k_B$. Being thorough, the entropy difference of $^{196}$Pt and $^{198}$Au corresponds to the sum of the proton and neutron contributions: $\Delta S\approx 2.2~k_B\approx\Delta S^{p} +\Delta S^{n}$.

In the microcanonical ensemble, the temperature of the system can be derived by the differentiation of the entropy with respect to the excitation energy:
\begin{equation}
\frac{1}{T(E)}=\frac{\partial S}{\partial E} 
 \label{eq-9}
\end{equation}
The statistical error and uncertainties from  the unfolding and the first generation method are propagated accordingly.
Since our experimental data do not constitute a continuous function and are affected by fluctuations, the derivative of $S(E)$ is obtained as a quadratic fit of every point with the four adjacent ones at the time. Therefore, the resulting temperature is smoothed over 0.6 MeV. Even though this procedure reduces the resolution below the experimental level ($\sim 0.15$ MeV), it is still possible to extract valuable information from the resulting caloric curves, $T(E)$, seen in Fig.~\ref{temperature_Pt_Au}.

For $^{197}$Au, large error bars and fluctuations are present, due to the relatively poor statistics of the dataset.
However, for all the nuclear systems under study the temperature increases up to 0.9 MeV and then drops to a minimum within a few hundreds keV. A decrease in temperature as a function of the excitation energy corresponds to the energy spent to break a pair of quasiparticles in the system. As one can notice, the most prominent peak is observed in the caloric curve of the only even-even nucleus presented in this work, $^{196}$Pt. The minimum is reached at 1.8 MeV in this case. This value is roughly equal to twice the pairing gap parameter $\Delta$. Soon after the temperature increases again, due to the latent pairing energy released to the isolated system. This characteristic fall and rise of the temperature can be clearly observed in the Pt isotopes between 1.0 and 2.2 MeV. Above this threshold the temperature reaches a value close to 0.63 MeV. The most disordered system, the odd-odd $^{198}$Au, manifests a very smooth entropy and, as a consequence, less structure in the temperature. This is visible in the quenching of the main bump at 0.9 MeV.

\begin{figure}[t]
\includegraphics[width=8.5cm,height=8.0cm]{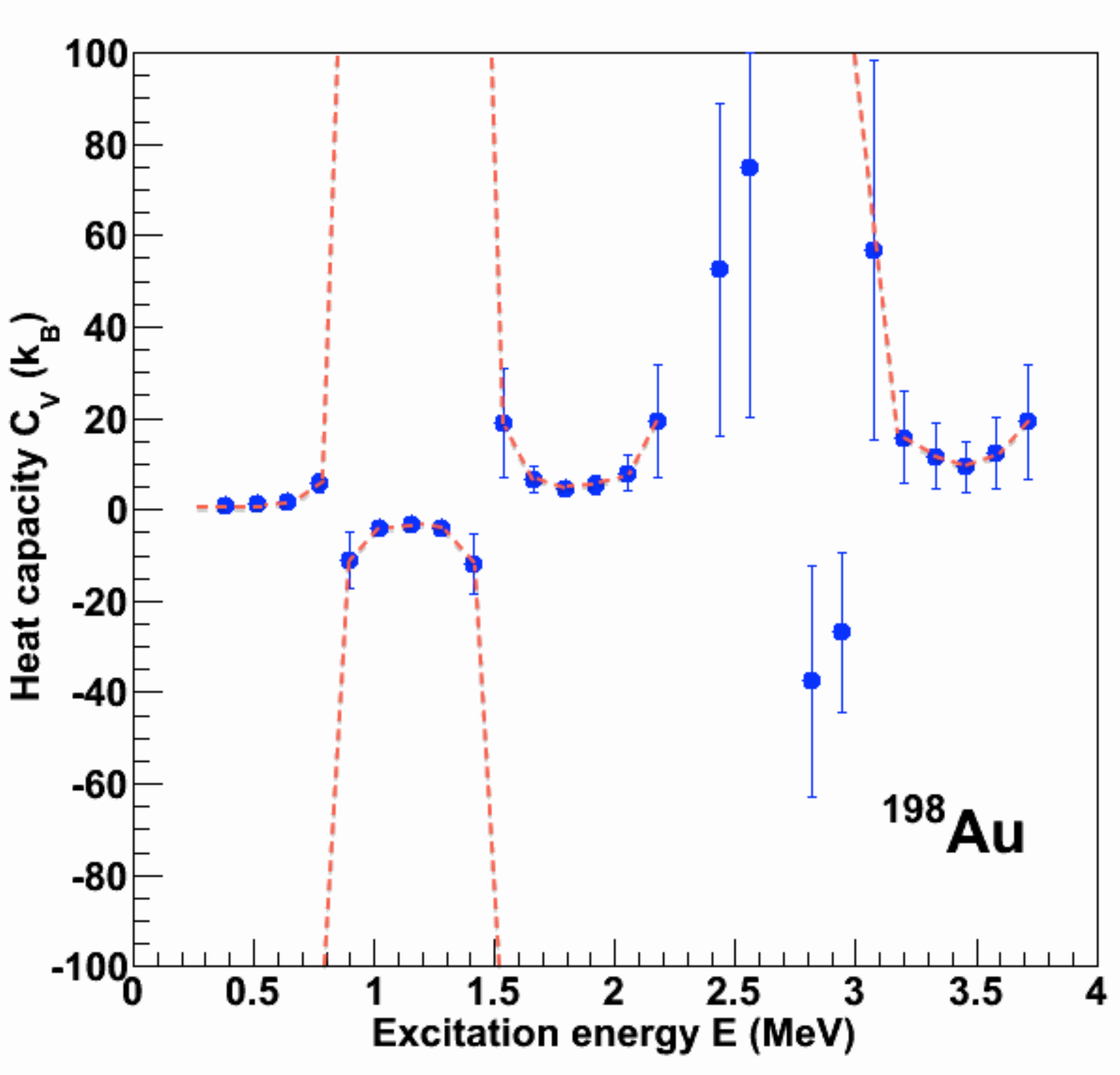}
\caption{(Color online) Microcanonical heat capacity of $^{198}$Au in the range of excitation energies between 0 and 4 MeV. The dashed red lines are drawn to help recognizing the sequence of the data points between divergences at $\pm\infty$.\label{heat_capacity_198Au}}
\end{figure}
 The even-even nuclear system, $^{196}$Pt, has the highest statistics and effective energy resolution. In this nucleus we can observe an oscillating distribution with secondary maxima at approximately 2.4, 3.4 and 4.2 MeV. Similarly to the most prominent bump at 0.9 MeV, we expect to observe other peaks at 4$\Delta$, 6$\Delta$ and so on, in correspondence of the energy threshold where two, three or more quasiparticle pairs can split. Nevertheless, the observed bumps occur after a spacing that is smaller than 2$\Delta$.

Moreover, pairing correlations are expected to become less important at higher excitation energies where the pairing field is decreased by the presence of a large number of unpaired quasiparticles. Further breaking of two or more Cooper pair is not noticeable in the level density and entropy distributions since the increase of the number of levels due to the splitting of a pair is smeared out. A linear increase in the entropy $S(E)$ is the overall effect (see Fig.~\ref{entropy_Pt_Au}), although small fluctuations around the mean value can still be observed in the caloric curve, as mentioned for $^{196}$Pt.

The peculiar fluctuating behavior of the temperature and in particular the decrease observed, for instance, between 1.0 and 1.8 MeV, corresponds to negative values of the heat capacity, derived as:
\begin{equation}
\frac{1}{C_{V}(E)}=\frac{\partial T}{\partial E}
 \label{eq-10}.
\end{equation}
This apparent violation of the laws of thermodynamics is related to the use of the microcanonical ensemble, which, in fact, is the most appropriate formalism to describe small systems. 
Negative heat capacities have been measured for several systems which are considered isolated, such as clusters of sodium atoms~\cite{SodiumCluster}, fragmenting Au nuclei~\cite{AuFragmentation} and magnetically self-confined plasma~\cite{plasma}. With the same analysis method presented in this work, negative heat capacity has been recently measured in actinides~\cite{actinidesMagne}. From the theoretical point of view, this feature has been subject of several interpretations. It has been addressed as a hint of first- or second-order phase transition~\cite{FirstOrder, SecondOrder}.

Recently it has been shown that atomic nuclei undergo a first-order phase transition from a superfluid to an ideal gas of non interacting quasiparticles being characterized by pairing and shell gaps in their particle spectra~\cite{MorettoPRL}: the heat capacity has an exponential dependence on the temperature up to when the critical point is reached and  the system has essentially absorbed all the energy. Similarly to a melting ice cube where the phase transition of water molecules is linearly dependent of the absorbed heat, quasiparticles are created with a constant energy cost and contribute with a constant amount of entropy to the disorder of the system. 

In Fig.~\ref{heat_capacity_198Au} the experimental heat capacity of $^{198}$Au is shown for excitation energy up to $E=4.0$~MeV. The heat capacity diverges to $\pm\infty$ when the temperature is constant, i.~e.~in correspondence to the maxima and minima. $C_V$ rapidly increases from zero to $+\infty$ at 0.9 MeV. Then it proceeds from $-\infty$ and increases up to $-0.2$ $k_B$ before to drop again to $-\infty$ close at 1.5 MeV.~ A new branch diverging at $+\infty$ covers the energy range between 1.5 and 2.2 MeV. If we compare the $C_V$ distribution with the corresponding caloric curve in Fig.~\ref{entropy_Pt_Au}, we find that the negative branches occur when the temperature decreases and the quasiparticle pair breaks up. We may conclude that a sequential melting of Cooper pairs occurs in the region of low excitation energies, as evident from the oscillating feature of the caloric curve and the negative branches of the heat capacity distribution shown in Fig.~\ref{entropy_Pt_Au} and Fig.~\ref{temperature_Pt_Au}.

\section{ Conclusions \label{conclusions}}
Excited states of $^{194-196}$Pt and $^{197,198}$Au up to the neutron separation energy were populated in ($p,p'\gamma$), ($p,d\gamma$), ($d,p\gamma$), ($d,d'\gamma$) and ($^{3}$He, $^{3}$He$'\gamma$) reactions.
From the measured $\gamma$-ray spectra, the level density of the five nuclear systems has been extracted.
Both the Pt and Au groups show a level density consistent with a constant-temperature description and characterized by the same temperature $T_{CT}=0.63$ and $0.67$~MeV, respectively.
The isotopes with an unpaired neutron, i.~e.~$^{195}$Pt and $^{198}$Au, are characterized by an increased density of levels with respect to the other systems with an even number of neutrons. This same effect is visible in the entropy distribution. The entropy difference $\Delta S =1.9$ $k_B$ in the $^{195,196}$Pt pair is comparable to the value obtained in well deformed nuclei, meaning that the unpaired valence neutron has a comparable degree of freedom in terms of available orbital space. The entropy excess due to an unpaired proton $\Delta S =1.1$ $k_B$ is also of the same order as in actinides. We can conclude that transitional Pt and Au isotopes show the same statistical features of mid-shell deformed heavy nuclei: the residual nuclear interaction is dominated by pairing correlations, while shell effects are not noticed in spite of the vicinity of the $Z=82$ shell closure.
 
Temperature and heat capacity have been deduced from the microcanonical ensemble formalism. Sequential bumps in the caloric curve are fingerprints of consecutive breaking of nucleon Cooper pairs in the heating nuclear system, showing a transition from an ordered to a disordered phase similar to the transition from superfluidity in liquids. 

\begin{acknowledgments}
We would like to thank E. A. Olsen, J. M\"{u}ller, A. Semchenkov, and J. Wikne at the Oslo Cyclotron Laboratory for providing stable and high-quality beams. 
\end{acknowledgments}

\end{document}